\documentclass{icrc29}

\usepackage{graphicx,amssymb,amsmath,times,lineno}

\begin{document}
\title[The Central Laser Facility...]{The Central Laser Facility at the Pierre Auger Observatory}

\author [F. Arqueros et al.] {F. Arqueros,
	J. Bellido, C. Covault, D. D'Urso, C. Di Giulio, P. Facal, B. Fick, F. Guarino, 
        \newauthor
	M. Malek, J.A.J. Matthews, J. Matthews, R. Meyhandan, M. Monasor, M. Mostafa,
        \newauthor
	P. Petrinca,  M.  Roberts, P. Sommers, P. Travnicek, L. Valore, V. Verzi, L. Wiencke 
	\newauthor
	for the Pierre Auger Collaboration.}
\presenter{Coordinator: (L. Wiencke wiencke@cosmic.utah.edu usa-malek-M-abs1-he15-poster)}

\maketitle
\begin{abstract}
The Central Laser Facility is located near the middle of the Pierre
Auger Observatory in Argentina.  It features a UV laser and optics
that direct a beam of calibrated pulsed light into the sky.  Light
scattered from this beam produces tracks in the Auger optical
detectors which normally record nitrogen fluorescence tracks from
cosmic ray air showers.  The Central Laser Facility provides a "test beam" to
investigate properties of the atmosphere and the fluorescence
detectors.  The laser can send light via optical fiber simultaneously
to the nearest surface detector tank for hybrid timing analyses.  We
describe the facility and show some examples of its many uses.
\end{abstract}

\section{Introduction}
The southern Pierre Auger Observatory in Mendoza Province, Argentina
measures extensive air-showers produced by cosmic rays.  A surface
detector (SD) records the distribution of charged particles at ground
level. A fluorescence detector (FD), consisting of independent eyes
operating at night, records light profiles of showers as they develop
in the atmosphere. The FD makes a calorimetric measurement of shower
energy because the amount of fluorescence light emitted is
proportional to the energy deposited in the atmosphere.

\begin{figure} 
 \centering 
 \includegraphics[width=0.65\textwidth]{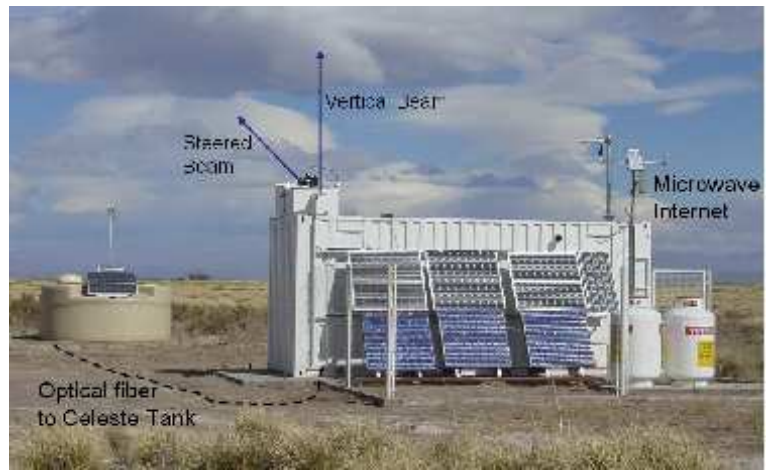}
 \caption{The Central Laser Facility at the Pierre Auger Southern Observatory.} 
 \label{clf-photo} 
\end{figure}

The Central Laser Facility (CLF) (Fig. \ref{clf-photo}) provides a
laser generated ``test beam'' for the observatory.  The CLF tracks
recorded by the FD eyes have several similarities to tracks generated
by extensive air showers.  The laser wavelength of 355 nm is near the
middle of the air shower fluorescence spectrum.  After accounting for
atmospheric effects, the total amount of light scattered out of the
beam is proportional to the laser beam energy.  At full laser energy,
the amount of light scattered out of the beam is roughly equivalent to
the amount of fluorescence light emitted by showers with energy in
region of sensitivity of the FD to the predicted GZK suppression.  The
CLF can direct light simultaneously into a SD tank via optical fiber
to test the relative timing between the FD and SD.  The facility also
houses a weather station and a radiometric cloud detector
\cite{R2004}.

\begin{figure*} [t]
\centering
\includegraphics[width=12.0cm]{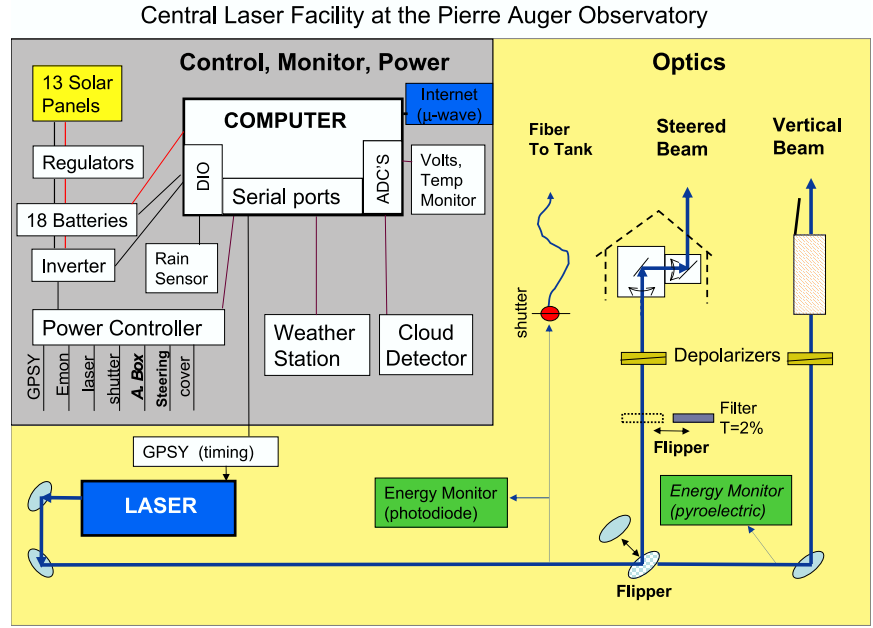}
\caption{The major components of the CLF.}
\label{clf-diagram} 
\end{figure*}

\section{Hardware Description}

The heart of the system is a frequency tripled YAG laser producing a 7
ns pulsed beam with a maximum energy per pulse of 7 mJ.  The optical
configuration (Fig. \ref{clf-diagram}) includes two harmonic separator
mirrors arranged so that the spectral purity of the 355 nm beam sent
to the sky is better than 99.9\%.  A ``flip mirror'' selects between
two vertical beam paths.  One goes directly to the sky when a simple
cover is open.  The other enters a mechanism with two mirrors on
rotating orthogonal axes that can steer the beam in any direction
above the horizon.  A mechanical cover protects the steering apparatus
when not in use.  The redundancy of this dual beam configuration
increases system reliability and simplifies operation.

The beams are calibrated and configured with randomized polarization.
The symmetric scattering properties of a depolarized beam are
desirable: in this case a vertical beam scatters light equally in the
direction of each FD eye.  Depolarizers in each beam path convolve the
linearly polarized beam emitted by the laser to an output beam for
which the net polarization is within 3\% of random.  The relative
energy of each pulse is monitored by a photo-diode detector.  A
pyroelectic energy probe, installed in March 2005, makes a second
relative measurement of the vertical beam energy.  This probe, factory
calibrated to NIST standards, is used to calibrate the photo-diode probe
in terms of energy delivered to the sky.  This procedure is performed
several times per year.  Energy and polarization calibrations are
performed downstream of the last optical element before the beams
enter the sky.

The CLF is solar powered, housed in a modified shipping container, and
operated remotely via microwave Internet link.  The facility is
unmanned.  No wires run to the CLF.  A small computer (PC) monitors
the solar power system, collects weather station and cloud detector
data, and operates the laser system.  All of the control software is
written using standard C programming tools.  The PC which consumes 5 watts,
runs a 25 MB version of embedded Linux, and has no moving parts.  Each
time the laser is fired, the time, relative energy, beam path ID, and
direction are recorded to a compact flash card.  All data are copied
off-site daily.

\begin{figure}
\begin{minipage}[t]{7.0cm}
\includegraphics*[width=1.0\textwidth]{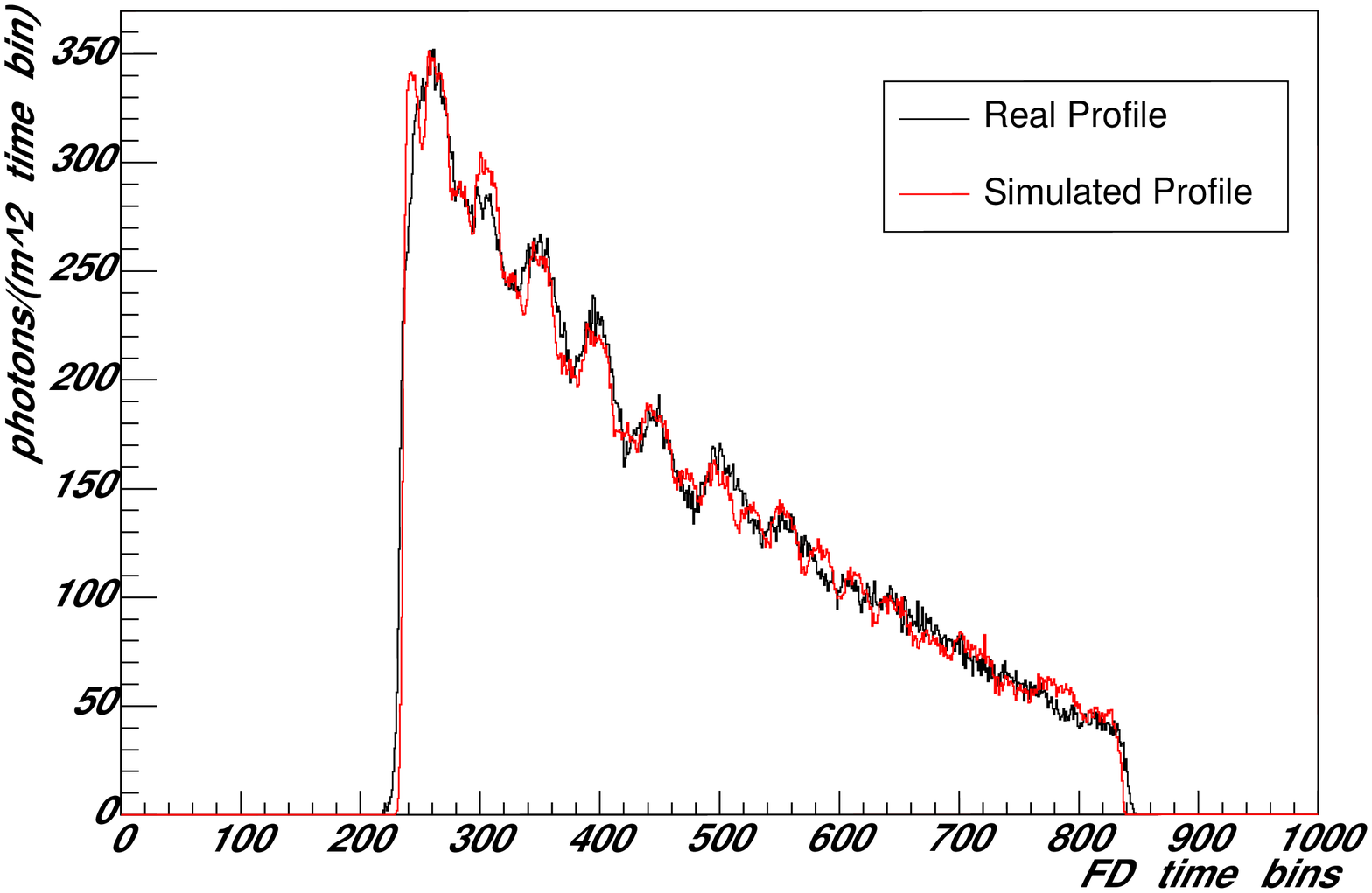}
\caption{\label{clf-pixel} Simulated and measured FD detector response
to a vertical CLF laser shot.  The detailed simulation in this example
includes a complete model of the detector optics, pixel elements, and
reflective triangular inserts between pixels. One FD time bin is 100 ns.}
\end{minipage}
\hfill
\begin{minipage}[t]{7.0cm}
\includegraphics*[width=1.0\textwidth]{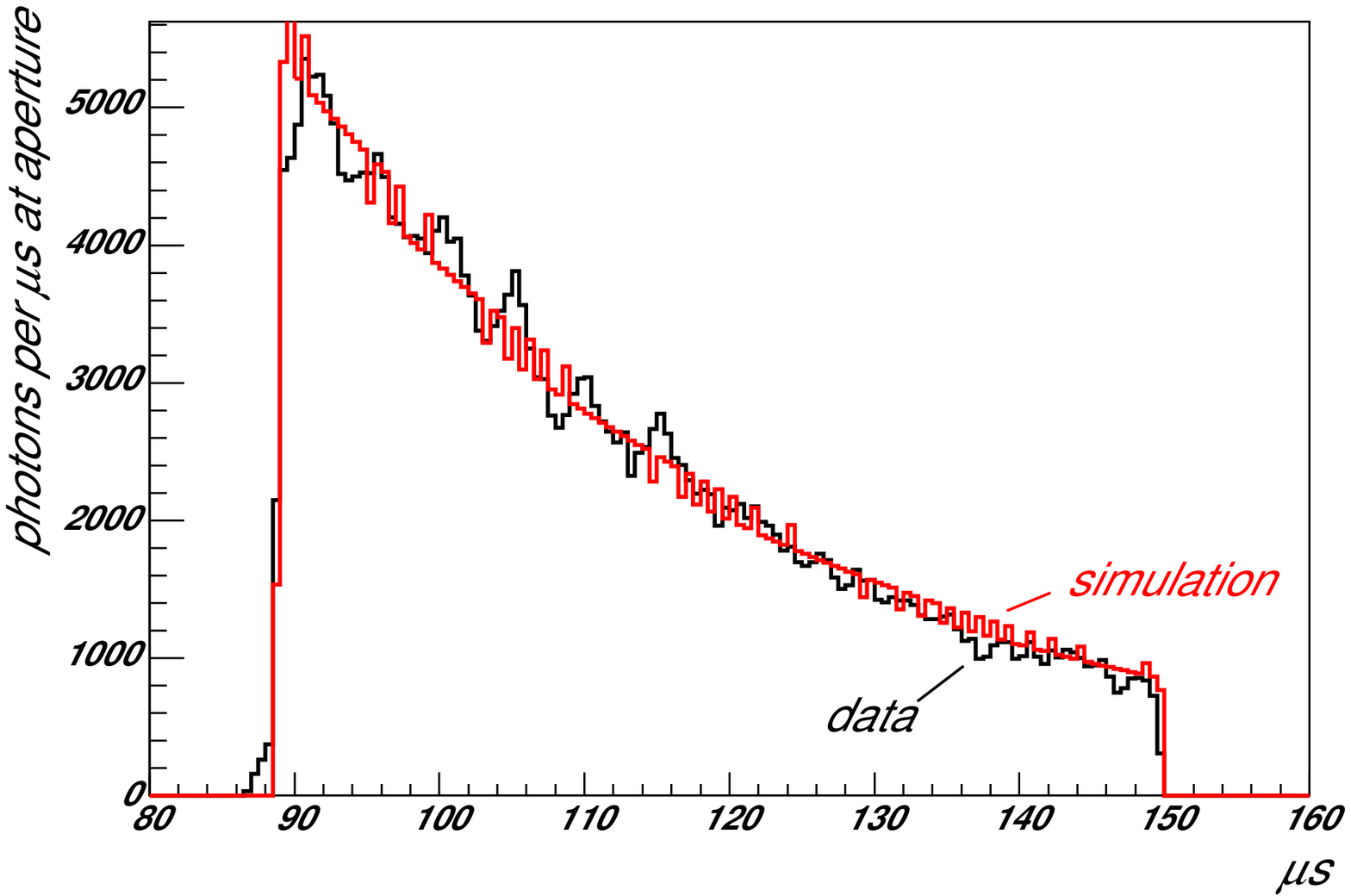}
\caption{\label{clf-profile}  Light profile at the aperture of the Los
Leones FD of a vertical CLF laser shot recorded under extremely clear
atmospheric conditions.  This simplified simulation assumes an 
aerosol-free atmosphere.  The horizontal axis corresponds to 
the time since the laser fired.}
\end{minipage}
\hfill
\end{figure}

The laser firing time relative to the GPS second is controlled by a
programmable timing module \cite{S2001}.  Every shot is tagged to
avoid confusion with possible neutrino-generated upward going tracks.
A small fraction of the laser light can be injected into the nearest
SD water tank (dubbed ``Celeste'') via optical fiber.  The ``sky-only''
shots are recorded by each triggered FD to a special laser file.  The
``sky+Celeste'' shots are written to the same data stream as air
shower candidates together with the corresponding data from the
Celeste tank.

Laser operation is conducted from the observatory's central campus by
starting and stopping the control program at the beginning and end of
each night of FD operation.  Sets of 50 fixed-direction vertical shots
are fired every 15 minutes followed by one sky+Celeste shot.  Patterns
of inclined shots are also fired during two one-hour periods each
night.  This pattern include shots fired along the bisector angles
between FDs and low-energy near-horizontal shots fired over each FD
eye.
\begin{figure} 
\centering 
\includegraphics[width=0.75\textwidth]{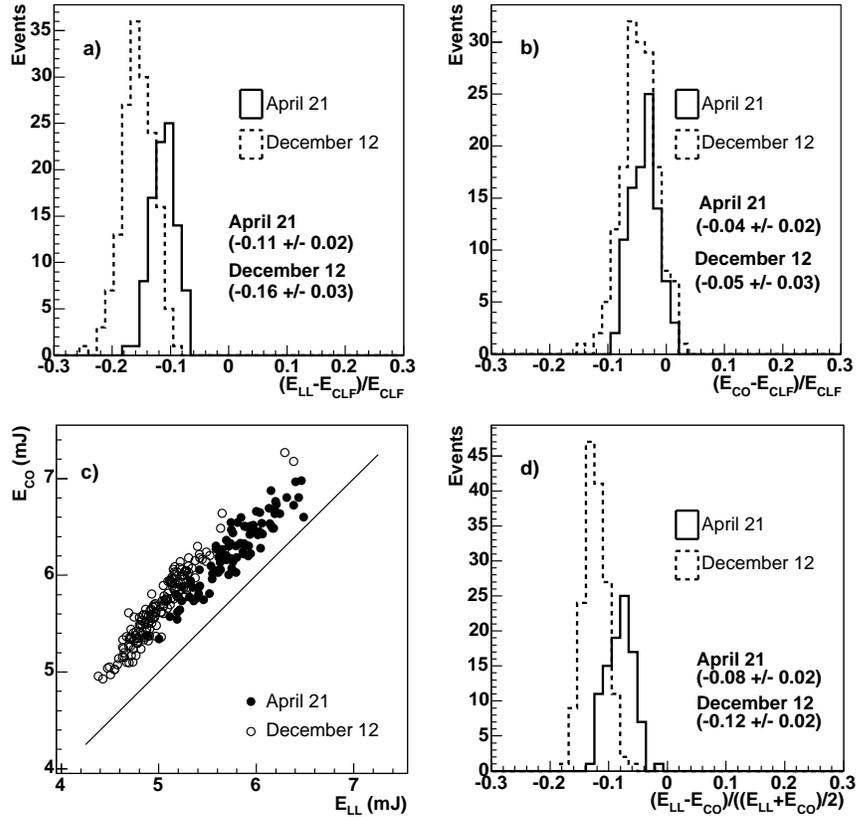}
\caption{A vertical beam from the CLF is used in a global cross-check
of photometric calibration for two nights with extremely clear viewing
conditions.  This cross-check involves two FD eyes, the CLF, the
molecular atmospheric description, and 60 km of total light path.  The
consistency in the measured and reconstructed CLF energies is at the
level of 15\% (a,b). The correlation in laser energy as reconstructed
by the Coihueco (CO) and Los Leones (LL) eyes (c,d) demonstrates the
sensitivity of this test.  The results are consistent with the current
level of uncertainty in the calibrations of the elements tested including
the atmosphere.}
\label{co-ll} 
\end{figure}

\section {Examples of Use}
On nights scheduled for FD operation, the observation of CLF tracks
provides real-time confirmation that the FD eyes are functioning and
are able to ``see'' the array center.  Shots fired into the sky and
the Celeste tank are used to monitor the relative timing between the
FD and the SD \cite{B2005}.  A fit to the longitudinal profile of
vertical tracks recorded by the FDs has been normalized to a clear air
profile to generate an hourly data base of aerosol optical depth
measurements \cite{R2005}.  CLF shots are used in ongoing studies
of triggering efficiency, geometrical reconstruction \cite{H2005} and the details
of the FD pixel-pixel response (Fig. \ref{clf-pixel}).  Sweeps of
inclined shots are also used to test the FD mirror pointing
directions.  Work in progress compares the profiles of CLF vertical
tracks recorded under extremely clear atmospheric conditions to
simulated profiles assuming an aerosol-free atmospheric model.
(Fig. \ref{clf-profile}).  These comparisons are used to cross-check
the calibration of the fluorescence telescopes
(Fig. \ref{co-ll}). This work tests, in combination, the photometric
calibration of the CLF and two FD eyes, the identification of clear
conditions, and modeling of molecular atmospheric conditions over a
total light path distance of 60 km.

\end{document}